\documentclass[conference]{IEEEtran}

\ifCLASSINFOpdf
\else
\fi

\usepackage{url}

\usepackage[]{fancyhdr} %
\newcommand{\changefont}{\fontsize{7}{7}\selectfont}
\usepackage[utf8]{inputenc}
\usepackage{amsmath,amssymb,amsfonts}
\usepackage{multicol}
\usepackage{multirow}
\usepackage{array}
\usepackage{graphicx}
\usepackage{cite}
\usepackage{mathtools}
\usepackage{booktabs}
\usepackage{xcolor}
\usepackage{hyperref}
\hypersetup{breaklinks=true}
\newcommand*\rot{\rotatebox{90}}

\IEEEoverridecommandlockouts
\begin{document}
\graphicspath{{./fig/}}
\bstctlcite{bstctl:nodash}
\title{Enabling Undergrounding \\ of Long-Distance Transmission Lines\\ with Low Frequency AC Technology}%
\author{\IEEEauthorblockN{David K. Sehloff, Line A. Roald}
\IEEEauthorblockA{Dept. of Electrical \& Computer Engineering\\University of Wisconsin-Madison\\
Madison, USA\\
\{dsehloff, roald\}@wisc.edu}\vspace{-\baselineskip}\thanks{This work was supported in part by New York Power Authority (NYPA), in part by New York State Energy Research and Development Authority (NYSERDA), and in part by National Science Foundation Graduate Research Fellowship Program under Grant DGE-1747503.}}%
\maketitle%
\thispagestyle{fancy}%
\pagestyle{fancy}%
\begin{abstract}
With increasing prevalence of severe natural hazards and the ignition of wildfires by power lines, many power system planners are considering converting overhead lines to underground cables to mitigate risks related to these events. Systems with a large proportion of underground cables can bring challenges due to the capacitance and losses of the cables, factors which may limit the potential for hardening in critical areas. Low frequency AC (LFAC) transmission solves these problems, as lowering the frequency decreases both the effects of capacitance and the losses. This paper presents a tractable frequency-dependent model for underground cables and incorporates it into the open source optimal power flow tool \emph{VariableFrequencyOPF.jl}. The model and implementation are used to demonstrate the benefits of LFAC in a case study involving two multi-terminal cable upgrades with LFAC, including a comparison with HVDC. The results demonstrate the value of LFAC systems for both power flow control and reduction of losses.
\end{abstract}


\IEEEpeerreviewmaketitle

\section{Introduction}
Planning power systems for reliable operation under natural hazards is becoming more important as these hazards increase in frequency and severity. Hardening the system by converting overhead lines to underground cables is effective in mitigating susceptibility to wind, ice, and lightning, and risk of wildfire ignition \cite{Wang2016EnhancingPS}.
Because these upgrades are costly and often involve critical transmission corridors, it is crucial to understand their overall system impacts.

Despite their benefits, underground cables pose challenges for system operations.
Cables have a higher shunt capacitance than lines, limiting the feasible length of AC cables and changing the reactive power balance of the system, which can cause overvoltage problems.
In addition, challenges associated with cooling of cables require strict thermal limits \cite{Zaborszky1954,weedy1980}.
Together, the significant reactive power flow and tighter thermal limits associated with underground cables may reduce active power capacity relative to the original overhead line, thereby increasing congestion in the system.
This means that undergrounding may not be feasible for critical lines that carry power across long distances, which significantly limits the potential for hardening by undergrounding.
One way of mitigating these problems is to operate cables with high voltage direct current (HVDC), which eliminates reactive power and can allow an increase in active power flow. The protection systems for DC current remain a challenge, however, particularly for multi-terminal systems.

A solution which addresses these limitations is the emerging application of low frequency AC (LFAC) systems. An LFAC system uses AC/AC converters to lower the frequency on one or more critical cables. Lowering the frequency decreases the shunt susceptance, which reduces the reactive power flow generated by the shunt elements. This in turn enables higher active power flow and longer cables \cite{Funaki2000}. In addition, the losses in cables due to current distribution in the conductor, induced currents in the metallic sheath, and leakage in the dielectric insulation can be large in long 50 or 60 Hz cables. These losses all exhibit strong dependence on frequency, decreasing as frequency decreases. Furthermore, apart from reducing the frequency, the AC/AC converters also provide active and reactive power flow control that can be used to route power more efficiently through the system.

Low frequency AC transmission systems for underground or undersea cables have been analyzed from several perspectives. The modular multilevel converter design has been considered in \cite{Miura2013,Carrasco2014,Liu2016}, and \cite{SorianoRangel2018} considered the effect of LFAC frequency on the design requirements of the converter. At a system level, \cite{Meere2017} studied the installation and operating costs of low frequency offshore wind collection networks, introducing an optimization over frequency to minimize the cost of an offshore wind system. The converters for low frequency or HVDC networks introduce significant cost, but this cost is shown to be outweighed by the benefits in certain situations.

Effectively analyzing the system-level feasibility of LFAC for specific hardening upgrades requires a quantification of costs and benefits based on the power flow capabilities and constraints of the system. This paper presents an optimization framework that computes an optimal economic dispatch with the cables and LFAC in place. We build on the AC optimal power flow (OPF) model presented in \cite{Sehloff2021}, which includes frequency as a variable. With this LFAC OPF model, system operators and utilities can analyze the system-level advantages of using LFAC for system hardening. However, the existing LFAC model in \cite{Sehloff2021} is focused on overhead lines. In overhead lines, the frequency dependence due to the current redistribution from the skin effect and proximity effects is small and often neglected in steady-state. In underground cables, however, these effects are stronger due to the induced currents in the metallic sheaths of the cables and the close spacing of adjacent conductors \cite{wedepohl1973transient}. In this paper, we therefore extend the modeling framework to include %
an accurate model of cables for non-standard frequencies.

The modeling of underground cables is well established, including the electrical parameters and current limits accounting for skin and proximity effects for standard transmission frequencies \cite{Carson1926,Wagner1933,Zaborszky1954,weedy1979,weedy1980}. %
However, the assumption of a standard frequency prevents their application to analysis of cable operation for LFAC. A method using a frequency-dependent correction term for resistance was proposed in \cite{WajeAndreassen2016}, but has not been fully validated and lacks consideration of other parameters. More accurate models that incorporate detailed frequency-dependence of the sequence impedance are used for analysis of transients with components at multiple frequencies \cite{wedepohl1973transient,de2009accuracy,dommel1986}, but these models involve complex functions which are intractable for optimal power flow problems.
To enable the accurate analysis of LFAC cables within an optimal power flow framework, we propose a new frequency-dependent model for power flow in cables. This is incorporated into the modeling framework in \cite{Sehloff2021}, and used to analyze system behavior when a large number of lines are undergrounded to reduce wildfire risk.

In summary, the contributions of this paper are 1) an approximate frequency-dependent cable model which is tractable for optimal power flow, 2) a variable frequency AC optimal power flow formulation which incorporates this model, and 3) application of the model to a scenario where overhead lines are undergrounded to reduce wildfire risk, including a benchmark against standard frequency or DC cables.%

The paper is organized as follows: Section \ref{sec:model} presents the approximate model, Section \ref{sec:powerflow} incorporates the model into power flow and optimal power flow, Section \ref{sec:casestudy} demonstrates the model in a case study, and finally Section \ref{sec:conclusion} concludes.

\section{Frequency Dependent Steady State Model for Underground Cables}\label{sec:model}
\subsection{Circuit Model}
Cable modeling begins with the complex phasor quantities for voltage $\underline{V}$ and current $\underline{I}$, which include the central conductor and conductive sheath for three phases, so that $\underline{V},\underline{I}\in\mathbb{C}^6$. Similar to the single-conductor problem, the multi-conductor Telegrapher's Equations describe the relationship between $\underline{V}$ and $\underline{I}$ at a point $x$ \cite{wedepohl1973transient},
\begin{equation}
    \frac{d}{dx}\underline{V}(x)=-Z(\omega)\underline{I}(x);\quad
    \frac{d}{dx}\underline{I}(x)=-Y(\omega)\underline{V}(x)\label{eq:tel},
\end{equation}
where $Z(\omega),Y(\omega)\in\mathbb{C}^{6\times 6}$ are the distributed series impedance and shunt admittance matrices per unit length, respectively, $Z(\omega)$ describing the frequency-dependent self and mutual impedance and $Y(\omega)$ representing the self and mutual admittance of each conductor. In the following discussion we write these frequency-dependent matrices as $Z$ and $Y$, for simplicity.

\subsection{Impedance and Admittance Model}\label{sec:zy}
\begin{figure}[!t]
\centering
\includegraphics[width=0.48\textwidth]{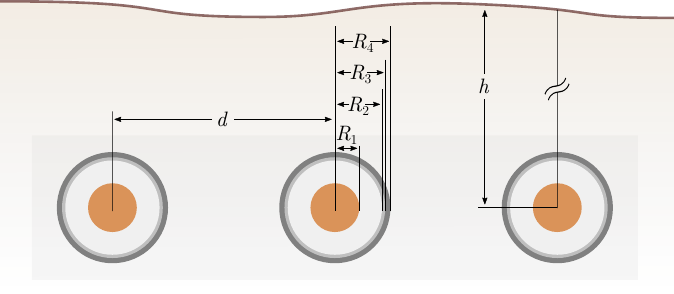}
\caption[Cross section of a three phase cable system in flat configuration.]{Cross section of a three phase cable system in flat configuration with the radii $R_1$-$R_4$, spacing $d$, and depth $h$ labeled. The radius of the central conductor is $R_1$, $R_2$ is the radius to the outside of the inner insulation, $R_3$ is the radius to the outside of the sheath, and $R_4$ is the outer radius.}
\label{fig:cable-diag}
\end{figure}

A variety of models and approximations have been proposed to construct the matrices $Z$ and $Y$ for underground cable systems, including detailed representations accounting for skin and proximity effects \cite{wedepohl1973transient,Carson1926,weedy1980,dommel1986}. Ref \cite{wedepohl1973transient} gave a detailed model accounting for skin and proximity effects in ten impedance and admittance parameters, as described in Table \ref{tab:cable-zy}, along with the cable properties used in their calculation. Here $\rho$ is electric resistivity, $\mu$ is magnetic permeability, and $\epsilon$ is dielectric permittivity. The cable system dimensions, depicted in Fig. \ref{fig:cable-diag} are as follows: $R_1$ is the radius of the central conductor, which may be copper or aluminum, $R_2$ is the radius from the cable center to the outside of the inner insulation (commonly cross-linked polyethylene, or XLPE), $R_3$ is the radius to the outside of the metallic sheath (often aluminum), $R_4$ is the radius to the outside of the outer insulation (XLPE), and $d$ is the distance between centers of adjacent cables. These parameters are readily available from cable manufacturers, as in \cite{tfkable}, and the material property values are well-known. This data, along with the cable depth and spacing, provide all the information needed for calculation of $Z$ and $Y$.

The matrices $Z$ and $Y$ depend on the internal impedances of the central conductor, sheath, and adjacent soil, the impedances due to time-varying magnetic fields in the inner and outer insulations, the impedances due to the induced voltages in the sheath due to the flows of current through the central conductor and the adjacent soil, and finally the mutual impedance between separate cables, each as a function of the cable dimensions and materials and the electrical frequency. We follow the procedure of \cite{wedepohl1973transient}, using the exact formulas to construct $Z$.

\begin{table}[!t]
    \caption{Components of the impedance and admittance matrices $Z$ and $Y$ from \cite{wedepohl1973transient} and the basic cable properties on which they depend.}
    \centering
\begin{tabular}{@{}lll@{}}
\toprule
 & source of impedance or admittance & parameter dependence \\ \midrule
$z_1$ & central conductor internal impedance & $\rho,\mu,\omega,R_1$\\
$z_2$ & inner insulation time-varying magnetic field & $\mu,\omega,R_1,R_2$\\
$z_3$ & inner sheath internal impedance & $\rho,\mu,\omega,R_2,R_3$\\
$z_4$ & sheath mutual impedance & $\rho,\mu,\omega,R_2,R_3$\\
$z_5$ & outer sheath internal impedance & $\rho,\mu,\omega,R_2,R_3$\\
$z_6$ & outer insulation time-varying magnetic field & $\mu,\omega,R_3,R_4$\\
$z_7$ & earth return impedance & $\mu,\omega,\rho,R_4,h$\\
$z_{ij}$ & inter-cable mutual impedance & $\mu,\omega,\rho,d,h$\\
$y_1$ & inner insulation leakage admittance & $\rho,\epsilon,\omega,R_1,R_2$\\
$y_2$ & outer insulation leakage admittance & $\rho,\epsilon,\omega,R_3,R_4$\\
 \bottomrule
\end{tabular}
\label{tab:cable-zy}
\end{table}

In addition, we follow \cite{wedepohl1973transient} to construct $Y$, modeling the leakage conductances and admittances between the central conductor and sheath and between the sheath and ground as functions of cable parameters and frequency. These electrostatic interactions are considered to be negligible between separate cables, so the inter-cable elements of $Y$ are zero.

\subsection{Calculating Positive Sequence Impedance}
The solution to the single-conductor version of (\ref{eq:tel}) is well-known. Similarly, the six-conductor problem has the following solution for a line of length $\ell$ with origin and destination voltages $\underline{V}_o$ and $\underline{V}_d$ and currents $\underline{I}_o$ and $\underline{I}_d$:
\begin{equation}
    \begin{bmatrix}
    \underline{V}_o\\
    \underline{I}_o
    \end{bmatrix} = \begin{bmatrix}
     B& C(ZY)^{-1/2}Z\\
    Z^{-1}(ZY)^{1/2}C & Z^{-1}BZ
    \end{bmatrix}
    \begin{bmatrix}
    \underline{V}_d\\
    \underline{I}_d
    \end{bmatrix}\label{eq:tel_sol},
\end{equation}
where $B\!=\!\frac{1}{2}\!\left(\!e^{\ell\sqrt{ZY}}\!+\!e^{-\ell\sqrt{ZY}}\!\right)\!$ and $C\!=\!\frac{1}{2}\!\left(\!e^{\ell\sqrt{ZY}}\!-\!e^{-\ell\sqrt{ZY}}\!\right)\!$.

We rewrite (\ref{eq:tel_sol}), partitioning the matrix to separate the three phase voltages and currents of the central conductors, $\underline{V}^{abc}$, $\underline{I}^{abc}$ and the sheaths, $\underline{V}^{\mathrm{s},abc}$, $\underline{I}^{\mathrm{s},abc}$:
\begin{equation}
    \begin{bmatrix}
    \underline{V}_o^{abc}\\
    \underline{V}_o^{\mathrm{s},abc}\\
    \underline{I}_o^{abc}\\
    \underline{I}_o^{\mathrm{s},abc}
    \end{bmatrix} = \left[\begin{array}{c|c|c|c}
    \alpha_{11} & \alpha_{12} & \alpha_{13} & \alpha_{14} \\\hline
    \alpha_{21} & \alpha_{22} & \alpha_{23} & \alpha_{24} \\\hline
    \alpha_{31} & \alpha_{32} & \alpha_{33} & \alpha_{34} \\\hline
    \alpha_{41} & \alpha_{42} & \alpha_{43} & \alpha_{44}
    \end{array}\right]
    \begin{bmatrix}
    \underline{V}_d^{abc}\\
    \underline{V}_d^{\mathrm{s},abc}\\
    \underline{I}_d^{abc}\\
    \underline{I}_d^{\mathrm{s},abc}
    \end{bmatrix},\label{eq:partitioned}
\end{equation}
where all partitions $\alpha_{11},\ldots,\alpha_{44}\in\mathbb{C}^{3\times3}$
 and can be computed as follows:
\begin{align}
\left[\begin{array}{cc}
    \alpha_{11} & \alpha_{12} \\
    \alpha_{21} & \alpha_{22}
    \end{array}\right] &=
    B\\
\left[\begin{array}{cc}
    \alpha_{13} & \alpha_{14} \\
    \alpha_{23} & \alpha_{24}
    \end{array}\right] &=
    C(ZY)^{-1/2}Z\\
\left[\begin{array}{cc}
    \alpha_{31} & \alpha_{32} \\
    \alpha_{41} & \alpha_{42}
    \end{array}\right] &=
    Z^{-1}(ZY)^{1/2}C\\
\left[\begin{array}{cc}
    \alpha_{33} & \alpha_{34} \\
    \alpha_{43} & \alpha_{44}
    \end{array}\right] &=
    Z^{-1}BZ
\end{align}

We assume the common practice of single point bonding \cite{Tziouvaras2006} is used for the cables, so that the sheaths are bonded together and to ground only at the origin side of the cable. Other methods such as cross bonding and solid bonding can be modeled similarly \cite{weedy1980}. Single point bonding at the origin side results in origin sheath voltages $\underline{V}_o^{\mathrm{s},abc}$ and destination sheath currents $\underline{I}_d^{\mathrm{s},abc}$ equal to zero.
Using this with \eqref{eq:partitioned} allows elimination of the sheath variables.
We then obtain a system of equations for the three phase voltages and currents:
\begin{equation}
    \begin{bmatrix}
    \!\underline{V}_o^{abc}\!\\
    \!\underline{I}_o^{abc}\!
    \end{bmatrix}\!=\!\left[\begin{array}{c|c}
    \alpha_{11}\!-\!\alpha_{12}\alpha_{22}^{-1}\alpha_{21}\!& \alpha_{13}\!-\!\alpha_{12}\alpha_{22}^{-1}\alpha_{23}\!\\\hline
    \alpha_{31}\!-\!\alpha_{32}\alpha_{22}^{-1}\alpha_{21}\!& \alpha_{33}\!-\!\alpha_{32}\alpha_{22}^{-1}\alpha_{23}\!
    \end{array}\right]\!
    \begin{bmatrix}
    \underline{V}_d^{abc}\\
    \underline{I}_d^{abc}
    \end{bmatrix}\label{eq:abc}.
\end{equation}
We transform this into the sequence components of the voltage and current, $\underline{V}^{012}$ and $\underline{I}^{012}$, using the Fortescue sequence transformation matrix $A$ \cite{kersting2006distribution}:
\begin{equation}
    \begin{bmatrix}
    \underline{V}_o^{012}\\
    \underline{I}_o^{012}
    \end{bmatrix} =
    \begin{bmatrix}
    A^{-1} & 0\\
    0 & A^{-1}
    \end{bmatrix}
    \begin{bmatrix}
    \underline{V}_o^{abc}\\
    \underline{I}_o^{abc}
    \end{bmatrix}\label{eq:seq}.
\end{equation}
Assuming a balanced, uniformly transposed system, we consider only the positive sequence voltages and currents, $\underline{V}^{1}$ and $\underline{I}^{1}$. We use (\ref{eq:abc}) and (\ref{eq:seq}) and take the positive sequence values:
\begin{equation}
    \begin{bmatrix}
    \underline{V}_o^{1}\\
    \underline{I}_o^{1}
    \end{bmatrix} =
    \left[\begin{array}{cc}
    a & b\\
    c & d
    \end{array}\right]
    \begin{bmatrix}
    \underline{V}_d^{1}\\
    \underline{I}_d^{1}
    \end{bmatrix}\label{eq:positive_seq}.
\end{equation}

From the lumped parameter $\Pi$ model circuit, we solve for series impedance $Z^\mathrm{ser}$ and shunt admittance $Y^\mathrm{sh}$ in terms of the values in (\ref{eq:positive_seq}):
\begin{align}
    Z^\mathrm{ser} = \frac{(d-1)(d+1)}{c}, \quad\quad
    \frac{Y^\mathrm{sh}}{2} = \frac{2c}{(d+1)}\label{eq:y_shunt}
\end{align}
The values $c$ and $d$ are found as
\begin{align}
    &c = A_{s1}
    \left[\alpha_{31}\!-\!\alpha_{32}\alpha_{22}^{-1}\alpha_{21}\right]
    A_{s2}\\
    &d = A_{s1}
    \left[\alpha_{33}\!-\!\alpha_{32}\alpha_{22}^{-1}\alpha_{23}\right]
    A_{s2}, \\
    &\text{with }     A_{s1} \!=\! \frac{1}{3}\left[
    1,e^{j2\pi/3}\!\!,e^{-j2\pi/3}\right]\!;\,
    A_{s2} \!=\! \left[
    1,
    e^{-j2\pi/3}\!\!,
    e^{j2\pi/3}
    \right]^\top \nonumber
\end{align}
These impedance and admittance values are the exact solution of (\ref{eq:tel}) for the terminal quantities at a given frequency.

\subsection{Approximate Lumped Parameter Model}
To accurately capture the frequency dependence of the impedance and admittance and incorporate them in power flow representations, we fit them with polynomial functions of frequency.
\subsubsection{Reference Data}
We begin with the basic properties of the cable system, described in Section \ref{sec:zy}. Given the range of frequencies $[\omega_{\min}, \omega_{\max}]$ for which we require accuracy we generate $n$ data points $(\omega_i, Z^\mathrm{ser}_i, Y^\mathrm{sh}_i)$ by identifying a vector of adequate samples of frequencies, $\boldsymbol{\omega}$, calculating $Z$ and $Y$ at each $\omega_i$ and solving (\ref{eq:y_shunt}) to obtain the values for $Z^\mathrm{ser}_i$, $Y^\mathrm{sh}_i$. These make up the vectors $\mathbf{Z^\mathrm{ser}}$ and $\mathbf{Y^\mathrm{sh}}$.

\subsubsection{Series Resistance}
For the series resistance we fit a quadratic model
\begin{equation}
    \tilde{R}(\omega)=r_2\omega^2+r_1\omega+r_0\label{eq:approx_r}
\end{equation}
to the reference data using least squares. Specifically, to obtain the parameters $r_2$, $r_1$ and $r_0$, we construct the matrix of independent variables $\mathbf{\Omega}_{r}$ from the reference frequencies,
    $\mathbf{\Omega}_{r} = [\boldsymbol{\omega}^2\quad \boldsymbol{\omega}\quad \mathbf{1}],$
where $\mathbf{1}$ is an $n\times 1$ vector of ones, and $\boldsymbol{\omega}^2$ represents an elementwise operation on the vector. We then calculate the coefficients:
\begin{equation}
    \begin{bmatrix}
    r_2&
    r_1&
    r_0
    \end{bmatrix}^\top = \left(\mathbf{\Omega}_{r}^\top\mathbf{\Omega}_{r}\right)^{-1}\mathbf{\Omega}_{r}^\top \operatorname{Re}\left(\mathbf{Z}^\mathrm{ser}\right).
\end{equation}

\subsubsection{Series Reactance and Shunt Susceptance}
We follow the same procedure for the series reactance and susceptance, but we set $x_0=0$ and $b_0=0$ to ensure that the values are zero at DC, giving us the quadratic models
\begin{align}
    \tilde{X}(\omega)&=x_2\omega^2+x_1\omega\label{eq:approx_x}\\
    \tilde{B}^\mathrm{sh}(\omega)&=b_2\omega^2+b_1\omega\label{eq:approx_b}.
\end{align}
We identify the coefficients $x_2$, $x_1$, $b_2$ and $b_1$ by defining
    $\mathbf{\Omega}_{x} = [\boldsymbol{\omega}^2\quad \boldsymbol{\omega}],$
and solving the least squares equations,
\begin{align}
    \begin{bmatrix}
    x_2&
    x_1
    \end{bmatrix}^\top = \left(\mathbf{\Omega}_{x}^\top\mathbf{\Omega}_{x}\right)^{-1}\mathbf{\Omega}_{x}^\top\operatorname{Im}\left(\mathbf{Z}^\mathrm{ser}\right)\\
     \begin{bmatrix}
    b_2&
    b_1
    \end{bmatrix}^\top = \left(\mathbf{\Omega}_{x}^\top\mathbf{\Omega}_{x}\right)^{-1}\mathbf{\Omega}_{x}^\top\operatorname{Im}\left(\mathbf{Y}^\mathrm{sh}\right).
\end{align}

\subsubsection{Shunt Conductance}
Finally, we use a quartic model for the shunt conductance, which exhibits large changes as frequency increases:
\begin{equation}
    \tilde{G}^\mathrm{sh}(\omega)=g_4\omega^4+g_3\omega^3+g_2\omega^2+g_1\omega+g_0\label{eq:approx_g}.
\end{equation}
We find the conductance coefficients $g_4$, $g_3$, $g_2$, $g_1$ and $g_0$ by defining
    $\mathbf{\Omega}_{g} = [\boldsymbol{\omega}^4\quad\boldsymbol{\omega}^3\quad\boldsymbol{\omega}^2\quad \boldsymbol{\omega}\quad \mathbf{1}],$
and solving
\begin{equation}
    \begin{bmatrix}
    g_4\!&\!
    g_3\!&\!
    g_2\!&\!
    g_1\!&\!
    g_0\!
    \end{bmatrix}^\top = \left(\mathbf{\Omega}_{g}^\top\mathbf{\Omega}_{g}\right)^{-1}\mathbf{\Omega}_{g}^\top\operatorname{Re}\left(\mathbf{Y}^\mathrm{sh}\right).
\end{equation}

\section{Optimal Power Flow with LFAC Cables}\label{sec:powerflow}
In this section, we present a model for power flow in systems with one or more LFAC branches. The model is largely a review of the model in \cite{Sehloff2021} but adds details on how to incorporate LFAC cables. In this framework, each portion of the network which has an independent frequency is called a \emph{subnetwork}, and each subnetwork can transfer power to another only through a \emph{frequency conversion interface}, which represents the AC/AC converter. The set of all subnetworks is $\mathcal{S}$, and the set of all frequency conversion interfaces is $\mathcal{I}$. The frequency of subnetwork $l$ is $\omega_l$. If a subnetwork has variable frequency, $\omega_l$ must remain within a range $\left[\underline{\omega_l},\overline{\omega_l}\right]$, determined by factors including the frequency converter capabilities, protection systems, and other devices in the subnetwork:
\begin{equation}
	\underline{\omega_l} \leq \omega_l \leq\overline{\omega_l},\qquad \forall l\in\mathcal{S}_\text{vf}\label{eq:var_f},
\end{equation}
where $\mathcal{S}_\text{vf}$ is the set of subnetworks with variable frequency. For other subnetworks, we fix the frequency to 50 or 60 Hz.

\subsection{Branch Parameters}
\subsubsection{Underground Cables}
The set of all branches in subnetwork $l$, each represented as two unidirectional edges, is $\mathcal{E}_l$.
We represent the series admittance values on a branch, where $ije$ is the index of branch number $e$ from bus $i$ to bus $j$ and $\tilde{R}_{ije}(\omega)$ and $\tilde{X}_{ije}(\omega)$ are the functions given by (\ref{eq:approx_r}),(\ref{eq:approx_x}) for this branch. We denote the set of all cables in subnetwork $l$ as $\mathcal{E}^\mathrm{cable}_l$.
\begin{align}
    G_{ije}&=\frac{\tilde{R}_{ije}(\omega_l)}{\tilde{R}_{ije}^2(\omega_l)+\tilde{X}_{ije}^2(\omega_l)},&\forall ije\in\mathcal{E}^\mathrm{cable}_l,l\in\mathcal{S},\label{eq:susceptance}\\
    B_{ije}&=\frac{-\tilde{X}_{ije}(\omega_l)}{\tilde{R}_{ije}^2(\omega_l)+\tilde{X}_{ije}^2(\omega_l)},&\forall ije\in\mathcal{E}^\mathrm{cable}_l,l\in\mathcal{S}\label{eq:conductance}.
\end{align}
Similarly, the approximate shunt conductance and susceptance apply directly to the cables in each subnetwork:
\begin{equation}
    G^\mathrm{sh}_{ije}\!=\!\tilde{G}_{ije}(\omega_l);\,\,
    B^\mathrm{sh}_{ije}\!=\!\tilde{B}_{ije}(\omega_l),\,\,\forall ije\!\in\mathcal{E}^\mathrm{cable}_l,l\!\in\mathcal{S}\label{eq:sh_conductance}
\end{equation}
\subsubsection{Overhead Lines}
Overhead lines which do not have significant second-order frequency dependence from skin and proximity effects, such as overhead lines and transformers, are modeled in terms of constant resistance $R$, capacitance $C$, and inductance $L$:
\begin{align}
		G_{ije} &= \frac{R_{ije}}{R_{ije}^2+\omega_{l}^2 L_{ije}^2},&\forall  ije\in\mathcal{E}_l\setminus\mathcal{E}^\mathrm{cable}_l, l\in\mathcal{S}\label{eq:line_susceptance},\\
        B_{ije} &= -\frac{\omega_{l} L_{ije}}{R_{ije}^2+\omega_{l}^2 L_{ije}^2},&\forall  ije\in\mathcal{E}_l\setminus\mathcal{E}^\mathrm{cable}_l, l\in\mathcal{S}\label{eq:line_conductance}.
\end{align}
\subsubsection{Transformers}
We assume that transformers are present only in the standard-frequency subnetworks, and we model each as an ideal frequency-independent voltage step and phase shift connected to a $\Pi$ branch as in (\ref{eq:line_susceptance}),(\ref{eq:line_conductance}). The off-nominal turns ratio is $\tau$ and the phase shift is $\phi$.

\subsection{Frequency Conversion Interfaces}
Each converter $m$ injects active and reactive power into the network at its terminals $i$ and $j$, denoted $p_{im}^{I}$, $p_{jm}^{I}$, $q_{im}^{I}$, and $q_{jm}^{I}$. The total losses in the converter are $p^\mathrm{loss}_{m}$. To guarantee active power balance through the converter, we enforce that these sum to zero:
\begin{equation}
	p_{im}^{I} \!+\! p_{jm}^{I} \!+\! p^\mathrm{loss}_{m}\!=\!0,\quad\forall m \in \mathcal{I}.\label{eq:interface_pbal}
\end{equation}
We assume the use of a back-to-back or matrix modular multilevel converter or other topology which allows independent control of reactive power injection at each terminal. The topology and semiconductor and filter ratings determine the achievable voltage, current, and apparent power. For a detailed discussion of these losses and limits, refer to \cite{Sehloff2021}. There the converter losses were shown for an LFAC system with several converters at various frequencies. Since the losses are small and do not vary significantly with frequency and because the detailed loss model increases the computational complexity of the OPF formulation, we neglect converter losses for the purposes of this analysis.

\subsection{AC Power Flow}
We denote the voltage magnitude and angle at bus $i$ as $V_i$ and $\theta_i$, respectively.
The AC power flow equations give active and reactive power $p_{ije}^{E}$ and $q_{ije}^{E}$ injected from bus $i$:
\begin{align}
	\textstyle
  \MoveEqLeft[4] p_{ije}^{E}=\frac{V_i^2}{\tau_{ije}^2}\!\left(G_{ije}\! + \!G_{ije}^\textrm{sh}\right)
  \!-\!\frac{V_i V_j}{\tau_{ije}}\Big(G_{ije}\cos(\!\theta_i\!-\!\theta_j\!-\phi_{ije}\!)\notag\\&+B_{ije}\sin(\!\theta_i\!-\!\theta_j\!-\!\phi_{ije}\!)\Big),\forall ije\in\mathcal{E}_l, l\in\mathcal{S},\label{eq:p_o}\\
	\textstyle
	\MoveEqLeft[4] q_{ije}^{E}=-\frac{V_i^2}{\tau_{ije}^2} \left(B_{ije}+B_{ije}^\textrm{sh}\right)
	-\frac{V_i V_j}{\tau_{ije}}\Big(G_{ije}\sin(\!\theta_i\!-\theta_j\!-\phi_{ije})\notag\\&-B_{ije}\cos(\!\theta_i\!-\theta_j\!-\phi_{ije}\!)\Big),\forall ije\in\mathcal{E}_l, l\in\mathcal{S}\label{eq:q_o}.
\end{align}
We represent the set of all edges with origin at bus $i$ in subnetwork $l$ as $\mathcal{E}_{l,i}$.
The set of all buses in subnetwork $l$ is $\mathcal{N}_l$. Each bus may also have one or more generators in $\mathcal{G}_{l,i}$; the set of all generators at bus $i$ in subnetwork $l$. The set of all frequency converters connected to bus $i$ is $\mathcal{I}_i^N$. The total active and reactive power load at bus $i$ is $p_i^L$ and $q_i^L$, respectively.
\begin{align}
\textstyle
  \MoveEqLeft[16] \sum_{\mathclap{g\in \mathcal{G}_{l,i}}}p_g^{G}+\sum_{\mathclap{m\in\mathcal{I}_i^N}}p_{im}^I-\sum_{\mathclap{ije \in \mathcal{E}_{l,i}}}p_{ije}^{E}- p_i^L - V_i^2G_i^\text{sh}\!=\!0,\notag\\&\forall i\in\mathcal{N}_l, l\in\mathcal{S},\label{eq:pbal}\\
  \MoveEqLeft[16] \sum_{\mathclap{g\in \mathcal{G}_{l,i}}}q_g^{G}+\sum_{\mathclap{m\in\mathcal{I}_i^N}}q_{im}^I-\sum_{\mathclap{ije \in \mathcal{E}_{l,i}}}q_{ije}^{E}- q_i^L + V_i^2B_i^\text{sh}\!=\!0,\notag\\&\forall i\in\mathcal{N}_l, l\in\mathcal{S}.\label{eq:qbal}
\end{align}
Here the shunt conductance at bus $i$ is constant-valued $G_i^\text{sh}$, and the susceptance of discrete bus shunt elements is
\begin{equation}
	B_{i}^\textrm{sh} = \omega_l C_{i}^\textrm{sh}; \text{ or }B_{i}^\textrm{sh} = \frac{-1}{\omega_l L_{i}^\textrm{sh}},\quad\forall  i\in\mathcal{N}_l, l\in\mathcal{S}.\label{eq:susceptance_ch}
\end{equation}

The reference buses in each subnetwork, making up the set $\mathcal{N}_\text{ref}$, serve as the voltage angle reference of zero:
\begin{equation}
	\theta_i = 0, \qquad\forall i\in\mathcal{N}_\text{ref}\label{eq:vref}.
\end{equation}

\subsection{Engineering Limits}
The power flow is constrained by the following engineering limits.

\noindent\underline{Bus voltage magnitude:}
The voltage magnitude at each bus must remain in the range $\left[\underline{V_i},\overline{V_i}\right]$:
\begin{equation}
	\underline{V_i}\leq V_i\leq\overline{V_i}, \qquad \forall i\in\mathcal{N}_l, l\in\mathcal{S}\label{eq:vlim}.
\end{equation}
\noindent\underline{Thermal capacity:}
Thermal limits, arising from conductor heating and ampacity ratings, give the maximum apparent power flow, $\overline{s_{ije}^{E}}$, at each edge in the network:
\begin{align}
	\left(p_{ije}^{E}\right)^2 + (q_{ije}^{E})^2 \leq \left(\overline{s_{ije}^{E}}\right)^2,\qquad\forall ije\in\mathcal{E}_l, l\in\mathcal{S}.\label{eq:s_lim_f}
\end{align}
\noindent\underline{Angle limits:}
To ensure a margin for transient stability, the absolute angle difference of connected buses must be less than the angle limit $\overline{\theta}$, commonly set close to $40^\circ$\cite{kundur1994}:
\begin{equation}
	-\overline{\theta}\leq\theta_i - \theta_j\leq\overline{\theta}, \qquad\forall ije\in\mathcal{E}_l, l\in\mathcal{S}.\label{eq:ang_lim}
\end{equation}

\noindent\underline{Generator limits:}
Each generator has upper and lower limits for active and reactive power, $\overline{p_g^{G}}$, $\overline{q_g^{G}}$ and $\underline{p_g^{G}}$, $\underline{q_g^{G}}$:
\begin{equation}
	\underline{p_g^{G}}\leq p_g^{G}\leq\overline{p_g^{G}};\quad
	\underline{q_g^{G}}\leq q_g^{G}\leq\overline{q_g^{G}},\qquad\forall g\in\mathcal{G}_l, l\in\mathcal{S}.\label{eq:gen_pq}
\end{equation}

\subsection{Objective Function}
We set the objective to minimize total generation cost, $c^\mathrm{gen}$, as the sum of the convex cost function $c_g(\cdot)$ for each generator $g$, which depends only on active power injection of the generator.
\begin{equation}
c^\mathrm{gen} = \sum_{l\in \mathcal{S}}\sum_{g\in \mathcal{G}_l}c_g\left(p_g^{G}\right).\label{eq:obj}
\end{equation}
\subsection{Optimal Power Flow with Frequency-Dependent Cable Parameters}
Given the above modeling considerations, we present an AC OPF problem with multiple, variable frequencies and frequency-dependent cable parameters:

\begin{gather}
\begin{aligned}
	c^{\min} = &\min\quad c^\mathrm{gen}&&(\ref{eq:obj})\\
	\mathrm{s.t.}\quad &\text{$\omega$-dependent cable parameters}&&(\ref{eq:susceptance}-\ref{eq:sh_conductance})\\
	&\text{physical constraints}&&(\ref{eq:line_susceptance}-\ref{eq:interface_pbal})\\
	&\text{engineering limits}&&(\ref{eq:var_f},\ref{eq:vlim}-\ref{eq:gen_pq})
\end{aligned} \label{eq:opf}\tag{$\star$}
\end{gather}

This formulation and the cable modeling presented in the previous section is implemented in the open source Julia language package \emph{VariableFrequencyOPF.jl}\cite{VariableFrequencyOPF} initially developed as part of \cite{Sehloff2021}.

\section{Case Study: Undergrounding Lines for \\ Wildfire Risk Reduction}\label{sec:casestudy}
We demonstrate this formulation in the analysis of potential undergrounding upgrades for the RTS-GMLC network \cite{rts-gmlc}. We first describe the network and several potential undergrounding upgrades, then validate the cable model and analyze the impact of the upgrades on the overall system.

\subsection{RTS-GMLC System}
The RTS-GMLC network represents a transmission system with three areas and long overhead transmission lines between each of the areas, with lengths between 77 km and 135 km. The inclusion of geographic data in this test system allows exact determination of line lengths, which are used as an input to the cable model. The inter-area transmission lines in the system are primarily operated at 230 kV, while much of the transmission within each area is at 138 kV. The standard RTS-GMLC data includes one HVDC line with unlimited capacity, no losses, and a power-dependent operating cost. Because the goal of this analysis is to examine the potential for LFAC or HVDC to alleviate constrained transmission and reduce losses, we remove this HVDC line for our analysis.

\subsection{Undergrounding for Wildfire Risk Reduction}
The RTS-GMLC system has been previously used in analysis of public safety power shut-offs, where overhead lines are de-energized to avoid wildfire ignitions
\cite{Rhodes2021}.
These power shut-offs cause wide spread power outages. Undergrounding of lines is widely regarded as a less disruptive alternative to power shut-offs, as underground cables are unlikely to cause ignitions. Specifically, the utility PG\&E in California has announced plans to spend \$15-30 billion to underground 10\% of its lines\cite{pgeunderground}.
Here, we examine the impact of undergrounding on system operation, and analyze whether LFAC technology could be beneficial. %

\begin{figure}[!t]
\centering
\includegraphics[width=0.48\textwidth]{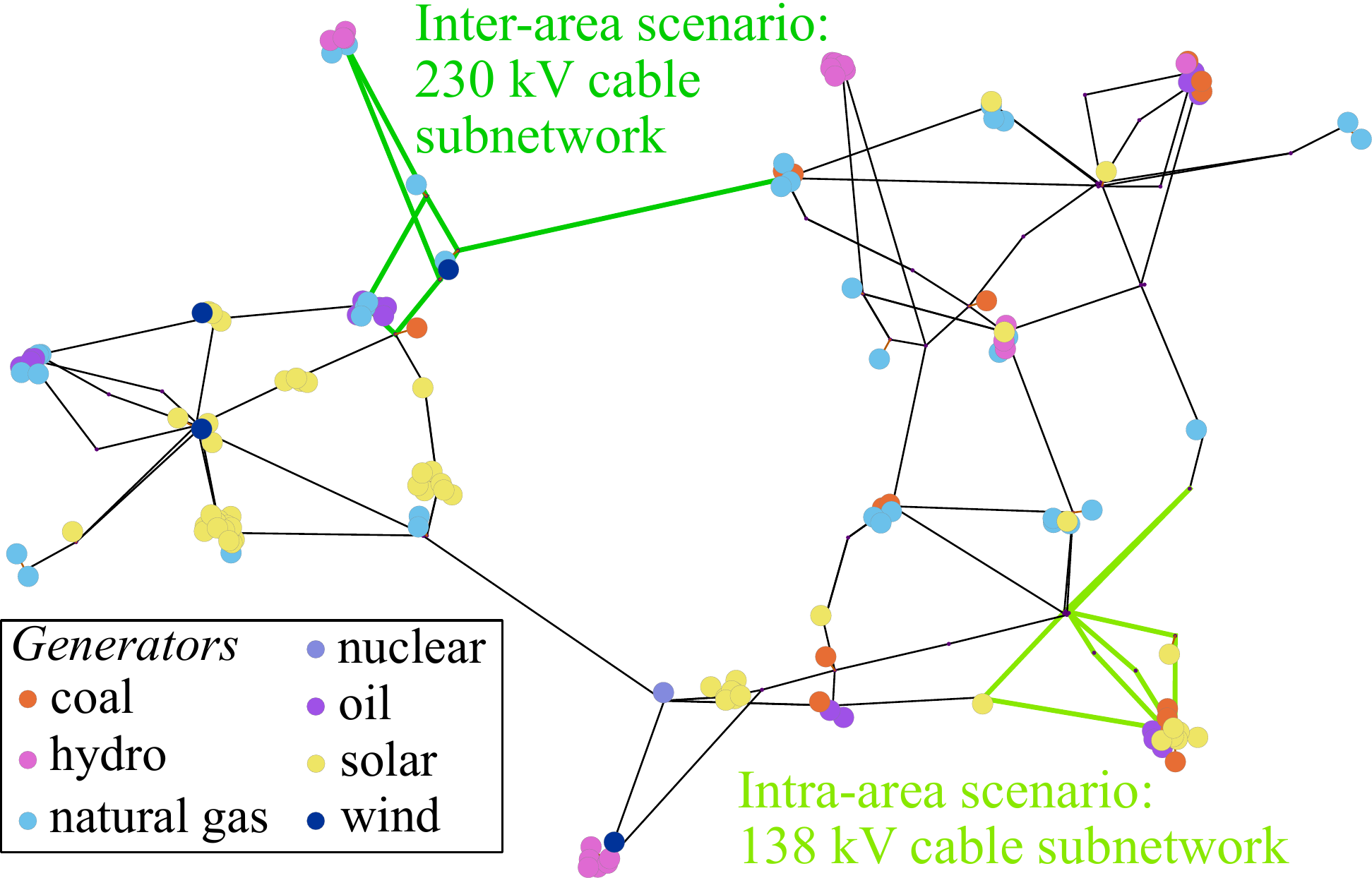}
\caption{The RTS-GMLC system, according to its geographic data. The branches involved in the two multi-terminal undergrounding scenarios are highlighted. This analysis considers each scenario separately.}
\label{fig:rts-diagram}
\end{figure}

For this case study, we utilize the results of \cite{Rhodes2021} to identify portions of the network with high risk of fire ignition, and we create two undergrounding upgrade scenarios. The first, the \emph{inter-area scenario},  involves the undergrounding of ten existing 230 kV overhead lines, all connected to form a multi-terminal cable network, including nine cables in one area, with lengths from 19 km to 124 km and one 135 km cable connecting them to another area. The second, the \emph{intra-area scenario}, is another multi-terminal undergrounding of ten existing overhead lines with lengths between 4 km and 54 km, this time all within a single area and having a voltage of 138 kV. Fig. \ref{fig:rts-diagram} illustrates the network and highlights the branches involved in the two scenarios, which we consider separately in the following analysis.

To convert the underground cables to LFAC, we add converters to connect the LFAC cable subnetwork to the main 60 Hz network and to connect loads and generators which continue to operate at 60 Hz. Table \ref{tab:converters} lists the buses at which the converters are added.

\begin{table}[!t]
    \caption{Buses at which converters are connected to apply LFAC to each scenario.}
    \label{tab:converters}
    \centering
\begin{tabular}{@{}lll@{}}
\toprule
         & inter-area scenario& intra-area scenario\\ \midrule
\shortstack{converters between LFAC\\and main 60 Hz network}  & 223,315,316 & 103,109,110\\\midrule
\shortstack{converters between LFAC\\and gens/loads} & 317,318,321,322 & 101,102,104,105,106\\ \bottomrule
\end{tabular}
\end{table}

\subsection{Parameters of Underground Cables}
For each overhead line that we plan to convert to an underground cable, we utilize a manufacturer datasheet \cite{tfkable} to select a cable which closely matches the voltage and thermal ratings of the existing overhead line. In scenario 1, the voltage rating must be at least 230 kV plus the allowed tolerance of 5\%, and the thermal limit of each of the lines is 500 MVA. We select a cable  rated for 245 kV and 525 MVA. In scenario 2, with 138 kV (+5\%) and 175 MVA line ratings, we select cables rated for 170 kV and 190 MVA. The properties of both of these cables, as taken from the datasheet, are shown in Table \ref{tab:cable-data}.
To calculate the detailed frequency-dependent matrices $Z$ and $Y$, we use constants for electric resistivity, magnetic permeability, and dielectric permittivity shown in Table \ref{tab:materials} for different cable materials.

\begin{table}%
    \caption{Parameters of the case study cables, selected from \cite{tfkable}.}
    \centering
\begin{tabular}{@{}lll@{}}
\toprule
                     & 230 kV Cable & 138 kV Cable \\ \midrule
voltage rating (kV)  & 245          & 170          \\
thermal rating (MVA) & 525          & 190          \\
$R_1$ (m)            & 0.0248       & 0.01515      \\
$R_2$ (m)            & 0.0479       & 0.03545      \\
$R_3$ (m)            & 0.0512       & 0.03835      \\
$R_4$ (m)            & 0.0565       & 0.043        \\
d (m)                & 0.226        & 0.172        \\
depth (m)            & 1            & 1            \\
conductor material   & Copper       & Copper       \\
sheath material      & Aluminum     & Aluminum     \\
insulation material  & XLPE         & XLPE         \\
configuration        & flat         & flat         \\ \bottomrule
\end{tabular}
\label{tab:cable-data}
\end{table}

\begin{table}%
    \caption{Material properties at $20^\circ$C for the detailed cable model.}
    \label{tab:materials}
    \centering
\begin{tabular}{@{}llll@{}}
\toprule
         & \shortstack{Resistivity\\$\rho$ ($\Omega\cdot m$)} & \shortstack{Permeability\\$\mu$ (H/m)} & \shortstack{Permittivity\\$\epsilon$ (F/m)} \\ \midrule
Copper   & $1.68\cdot10^{-8}$                                    & $1.25663\cdot10^{-6}$                         & -                                        \\
Aluminum & $2.65\cdot10^{-8}$                                    & $1.25667\cdot10^{-6}$                         & -                                        \\
XLPE     & $2.00\cdot10^{11}$                                    & -                                 & $2.3\cdot\epsilon_0$                           \\
Soil     & 100                                         & -                                 & -                                        \\ \bottomrule
\end{tabular}
\end{table}

We apply the modeling method from Section \ref{sec:model}, first solving for the detailed model parameters for each cable, then fitting an approximate model for impedance and admittance. For each fit, we set $\omega_{\min}=0.001$ and $\omega_{\max}=60\cdot2\pi$, and we use $n=500$ sampling points, for which the model can be computed and fit with no significant computational burden.

\subsection{Model Validation}
We first compare the detailed model to the manufacturer data and then compare the detailed and approximate model.
\subsubsection{Detailed Model Validation}
The cable manufacturer datasheet \cite{tfkable} gives values for distributed resistance per km at $90^\circ$C at 50 Hz. We solve for $Z^\mathrm{ser}$ and $Y^\mathrm{sh}$ in (\ref{eq:y_shunt}) for a 1 km cable using parameters for $90^\circ$C, and we obtain 50 Hz values of 0.0161 $\Omega$/km for the 230 kV cables and 0.0323 $\Omega$/km for the 138 kV cables. The datasheet gives 0.0182 $\Omega$/km and 0.0395 $\Omega$/km for these values, respectively, demonstrating agreement of the detailed model with empirical results.

\subsubsection{Approximate Model Validation}
\begin{figure}[!t]
\centering
\includegraphics[width=0.48\textwidth]{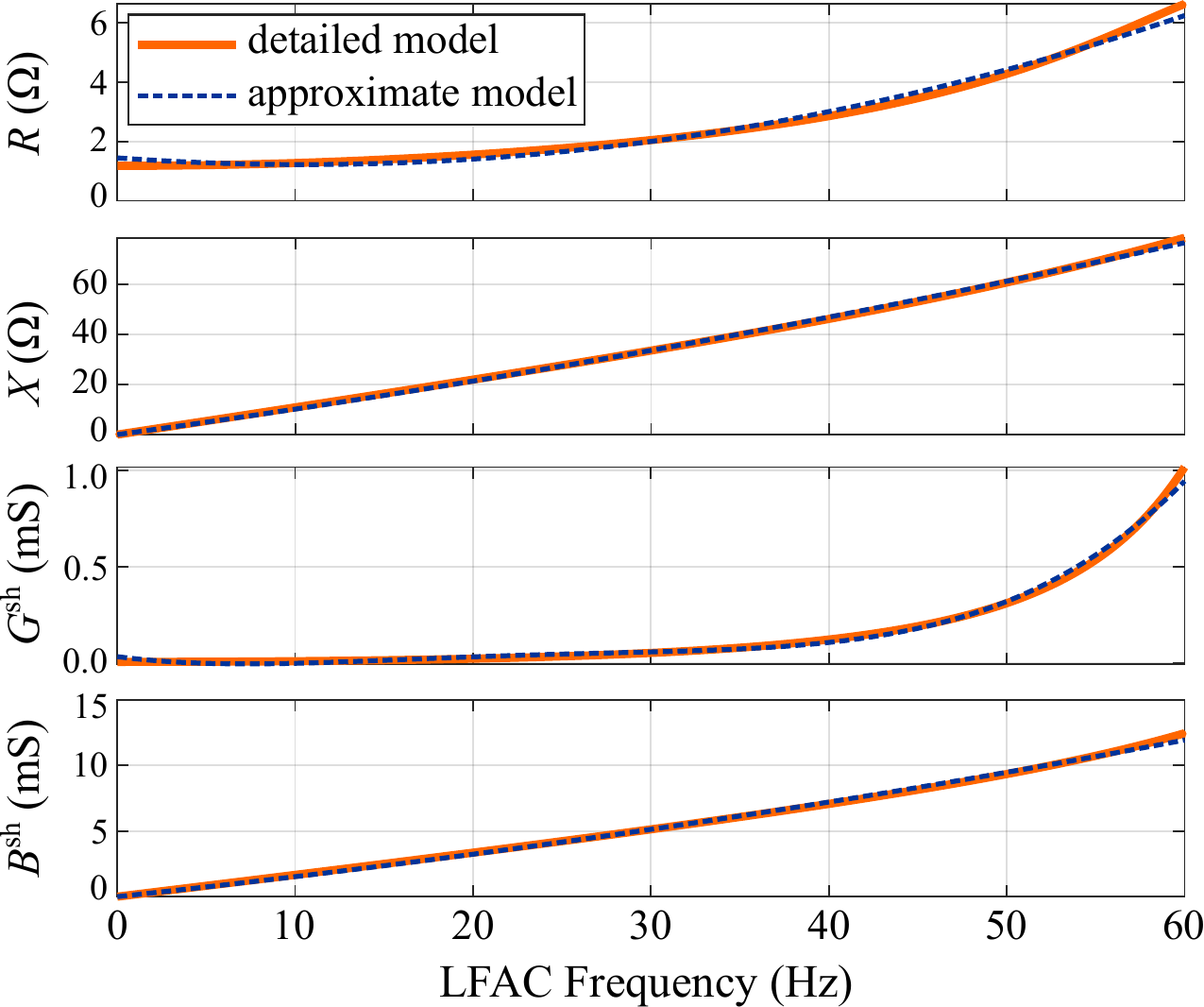}
\caption{Detailed and approximate frequency-dependent models for the 135 km, 230 kV cable in undergrounding scenario 1.}
\label{fig:fit}
\end{figure}

We find the approximate frequency dependent series resistance and reactance and shunt conductance and susceptance from (\ref{eq:approx_r}),(\ref{eq:approx_x}),(\ref{eq:approx_b}) and (\ref{eq:approx_g}) and compare them to the exact values from (\ref{eq:y_shunt}). Similar to the standard lumped parameter approximation for overhead lines, we observe higher errors for long lines due to the increased nonlinearity at long distances. The detailed and approximate model values for longest cable we consider, the 135 km, 230 kV cable, are shown in Fig. \ref{fig:fit}. Table \ref{tab:errors230} gives the largest error values, defined as the approximate value minus the detailed value, the relative error as a percentage of the largest value of the parameter, the frequencies at which these errors occur, and finally the RMS value of all errors for the cable. We include errors for both a 135 km, 230 kV inter-area cable and a 22 km, 138 kV intra-area cable. %

We observe that the largest errors occur at either 0 or 60 Hz. The largest errors for the 135 km cable is 7.1 \%, while the largest error for the 22 km cable is only 0.25 \%. The RMS errors are significantly smaller. We conclude that the proposed model is sufficiently accurate for our purposes.

\begin{table}[!t]
    \caption{Error of the approximate model compared to the detailed model for the 135 km, 230 kV and 22 km, 138 kV cables.}
    \label{tab:errors230}
    \centering
\begin{tabular}{@{}llllll@{}}
\toprule
 && \rot{\shortstack{Largest\\Error\\ Values ($\Omega$)}}     & \rot{\shortstack{Largest\\Relative\\ Error (\%)}} & \rot{\shortstack{Frequency \\ of Largest\\ Error (Hz)}}& \rot{\shortstack{RMS\\Error (\%)}}\\ \midrule
\multirow{4}{*}{\rot{\shortstack{135 km\\cable}}}& $\tilde{R}(\omega)-R(\omega)$ & -0.391 $\Omega$	    &	-5.9	&	60 & 2.2         \\
& $\tilde{X}(\omega)-X(\omega)$ & -1.98 $\Omega$	    &	-2.5	&	60 & 0.80        \\
& $\tilde{G}(\omega)-G(\omega)$ & -0.725 mS	&	-7.1	&	60 & 1.4          \\
& $\tilde{B}(\omega)-B(\omega)$ & -0.523 mS	&	-4.2	&	60 & 1.1         \\\midrule
\multirow{4}{*}{\rot{\shortstack{22 km\\cable}}}& $\tilde{R}(\omega)-R(\omega)$ & 1.55 m$\Omega$	    &	0.25	&	60 & 0.091         \\
& $\tilde{X}(\omega)-X(\omega)$ & -0.617 m$\Omega$	  &	-5.3$\cdot10^{-3}$ & 60 & 1.3$\cdot10^{-3}$\\
& $\tilde{G}(\omega)-G(\omega)$ &-12.2 pS  &-8.7$\cdot10^{-4}$ & 0 & 2.7$\cdot10^{-4}$\\
& $\tilde{B}(\omega)-B(\omega)$ & -0.224 $\mu$S	        &	-1.8$\cdot10^{-2}$	&	60 & 6.8$\cdot10^{-3}$         \\\bottomrule
\end{tabular}
\end{table}

\subsection{Power Flow Characteristics of a Single Cable}\label{sec:single-line}
\begin{figure}[!t]
\centering
\includegraphics[width=0.48\textwidth]{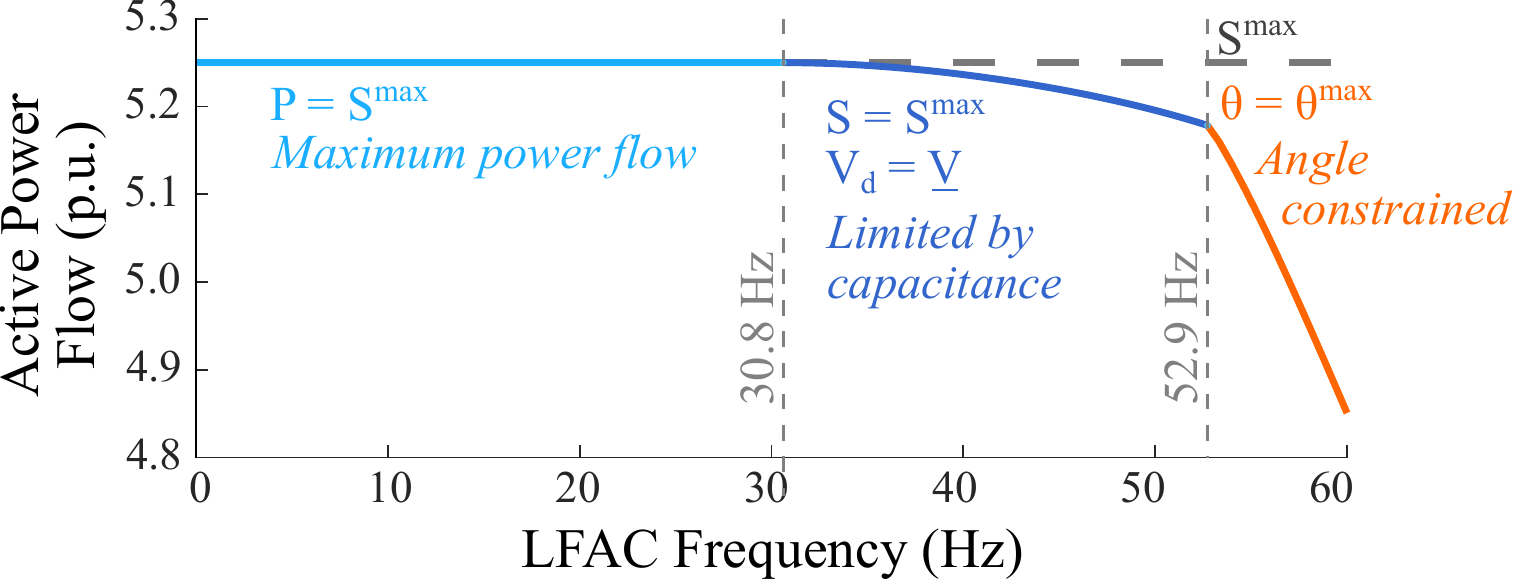}
\caption{Maximum power transfer on the 135 km, 230 kV cable as a function of frequency.}
\label{fig:pmax}
\end{figure}

We investigate the maximum power transfer on the 135 km, 230 kV LFAC cable. For this analysis,
we denote the origin and destination buses by $o$ and $d$, fix the destination voltage to 1.0 p.u. and set $\underline{V_d}=0.95$, $\overline{V_d}=1.05$, $\underline{\theta}=-40^{\circ}$, and $\overline{\theta}=40^{\circ}$. We then solve the single-cable optimization problem:
\begin{gather*}
\begin{aligned}
	\max\quad &P_o^E&&\\
	\mathrm{s.t.}\quad &\text{$\omega$-dependent power flow from $o$ to $d$}&&(\ref{eq:susceptance}-\ref{eq:q_o})\\
	&\text{engineering limits}&&(\ref{eq:vlim}-\ref{eq:ang_lim})\\
	&V_o = 1.0,\, \theta_o = 0&&
\end{aligned}
\end{gather*}
We solve this with a fixed frequency at increments from 0 and 60 Hz with results shown in Fig. \ref{fig:pmax}.

\subsubsection{Active Constraints}
In Fig. \ref{fig:pmax}, we observe three distinct frequency regions dominated by different constraints.

\noindent\underline{Angle Constrained:} From 60 Hz to 52.9 Hz, the active power is constrained by thermal limit (due to large reactive power flows) and the $40^\circ$ angle difference limit. The maximum power transfer capacity is very sensitive to the frequency. %

\noindent\underline{Capacitance Limited:} From 52.9 Hz to 30.8 Hz, the active power flow is constrained by the thermal limit and the reactive power generation of the cable.
The maximum active power is achieved when the destination voltage is at its lower limit, leading to the smallest levels of reactive power. The dependency on frequency is less pronounced, though still substantial.

\noindent\underline{Maximum Active Power:} From 30.8 Hz to 0 Hz, active power injection at the origin side matches the thermal limit. As a result, the maximum power transfer is not frequency dependent in this region. When the frequency reaches 0 Hz, power flow becomes a function of only voltage magnitude and resistance.

\subsubsection{Loss Characteristics} In the entire range from 60 to 0 Hz, the total loss on the cable decreases with frequency, with losses of 0.77 p.u. at 60 Hz dropping to less than 0.12 p.u at low frequencies. This can be explained both by the decreasing resistance and shunt conductance at low frequencies and the increase in voltage magnitude when the reactive power becomes less significant.

\subsection{Multi-Terminal Scenario Results}
Next, we analyze the two undergrounding scenarios.
\subsubsection{Base Case - No cables}
Solving the OPF for the base system without any underground cables or other upgrades, we find a minimum operating cost of 238.40$\cdot10^3$ USD. In this solution, the power loss in the network is 1.435 p.u.
\subsubsection{Inter-Area Scenario - Standard frequency cables}
We consider the inter-area scenario with undergrounding but no LFAC upgrades. We attempt to solve the OPF but converge to a point of infeasibility. Because this scenario involves the long, angle-constrained cable discussed in Section \ref{sec:single-line}, along with several other long cables, congestion and voltage issues introduced by the cables are likely factors in this infeasibility.

\subsubsection{Inter-Area Scenario - LFAC Cables and Converters}
Next, we apply the LFAC upgrades to the cable network and solve the OPF with the LFAC frequency as a variable, allowing it to vary between 0.1 Hz and 60 Hz. The optimal frequency in scenario 1 is 6.8 Hz, with a cost of 237.91$\cdot10^3$ USD. For comparison, we solve the OPF with the LFAC frequency at 60 Hz (standard frequency) and 16.7 Hz (a frequency used for railways, for which converters and switchgear are commercially available \cite{ABB2018}). In addition, we solve the OPF for an HVDC cable network, assuming a voltage rating $\sqrt{2}$ times larger than the AC rms value. We compare these results in the first column of Table \ref{tab:comparison}.

\begin{table}[!t]
    \caption{Comparison of the OPF solutions under each configuration.}
    \label{tab:comparison}
    \centering
\begin{tabular}{@{}rll@{}}
\toprule
                 & \multicolumn{2}{c}{cost ($10^3$ USD)}  \\ \cmidrule{2-3}
                  & scenario 1        & scenario 2         \\ \midrule
no LFAC           & infeasible        & 238.36             \\ \midrule
optimal frequency & 237.91 (@ 6.8 Hz) & 231.35 (@ 0.14 Hz) \\
16.7 Hz           & 238.04            & 231.36             \\
60 Hz             & 241.91            & 231.46             \\ \midrule
DC & 238.11            & 231.34             \\ \bottomrule
\end{tabular}
\end{table}

We note first that the addition of frequency converters allows a feasible solution to be found in all cases, even without using them to lower the frequency. This demonstrates to the value of active and reactive power control in networks with cables. Comparing the 60 Hz and optimal frequency solutions, we see a cost saving of about 1.7\% when the frequency is lowered. The solution at 16.7 Hz also achieves savings quite close to this. These solutions do not involve significant congestion, and the cables are operated well below their thermal limits, so the savings are largely due to reduced losses. The DC solution is also quite similar. At very low frequencies, we observe that voltage limits, both upper and lower, become active at several buses in the system, as power flow depends more strongly on voltage magnitude at low frequencies\cite{Ngo2016TD}. This may contribute to the slight increase in cost at very low frequencies and DC. Although the operational cost of DC is similar, multi-terminal DC networks bring a significant challenge, due to the difficulty of DC protection systems \cite{BLOND2016}.

\subsubsection{Inter-Area Scenario - Cost vs Frequency} 
\begin{figure}[!t]
\centering
\includegraphics[width=0.48\textwidth]{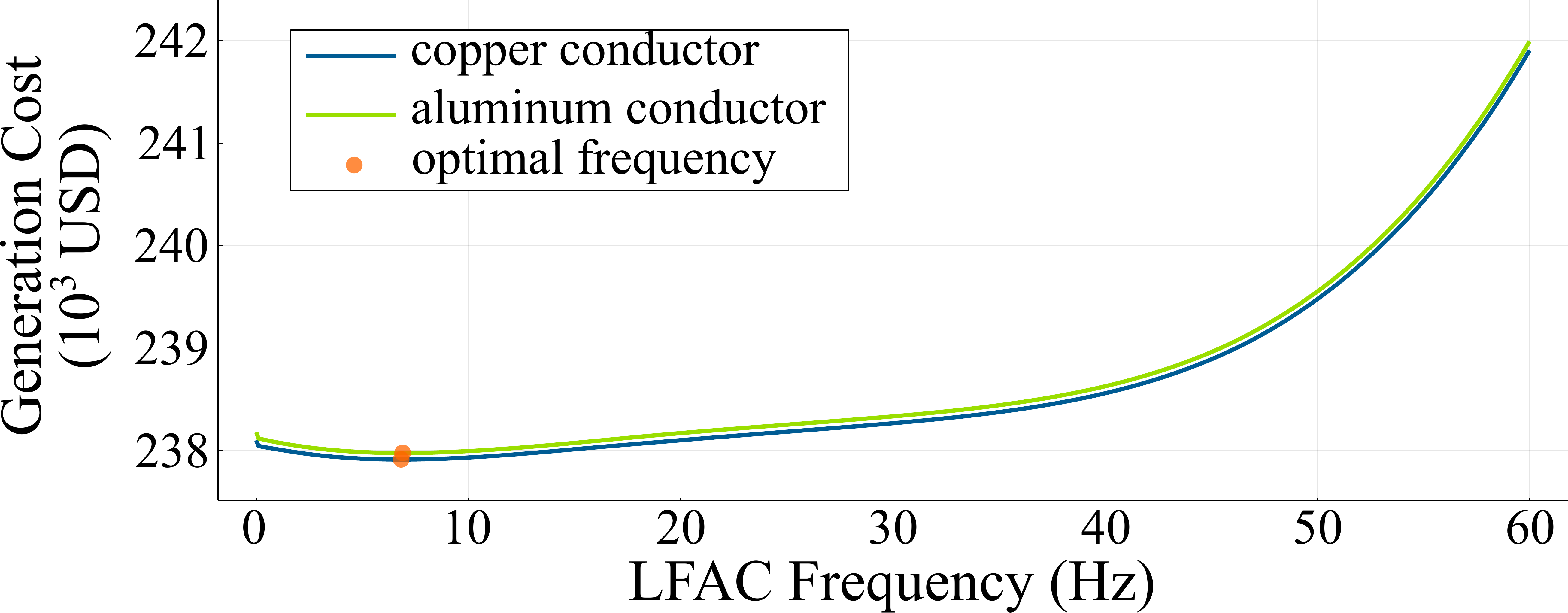}
\caption{Minimum generation cost for a sweep of LFAC frequencies in scenario 1. Cables with equal dimensions but aluminum conductors result in slightly higher cost than those with copper.}
\label{fig:sweepcost}
\end{figure}

We %
next perform a sweep of frequencies by fixing the LFAC frequency and solving the OPF at increments of 0.1 Hz from 0 Hz to 60 Hz. In addition, we model a cable with equivalent dimensions but an aluminum central conductor. This leads to slightly higher resistance and a lower thermal limit of 420 MVA. The resulting minimum cost for is shown in Fig. \ref{fig:sweepcost}. We first observe that the solution found for the variable frequency OPF problem is the lowest cost in the sweep and therefore not a local solution. Further, we observe a significant frequency dependence, especially close to 60 Hz. The difference between copper and aluminum is small; the lower thermal limit is not a factor because the cable network is not congested.

\subsubsection{Inter-Area Scenario - Losses vs Frequency} 
\begin{figure}[!t]
\centering
\includegraphics[width=0.48\textwidth]{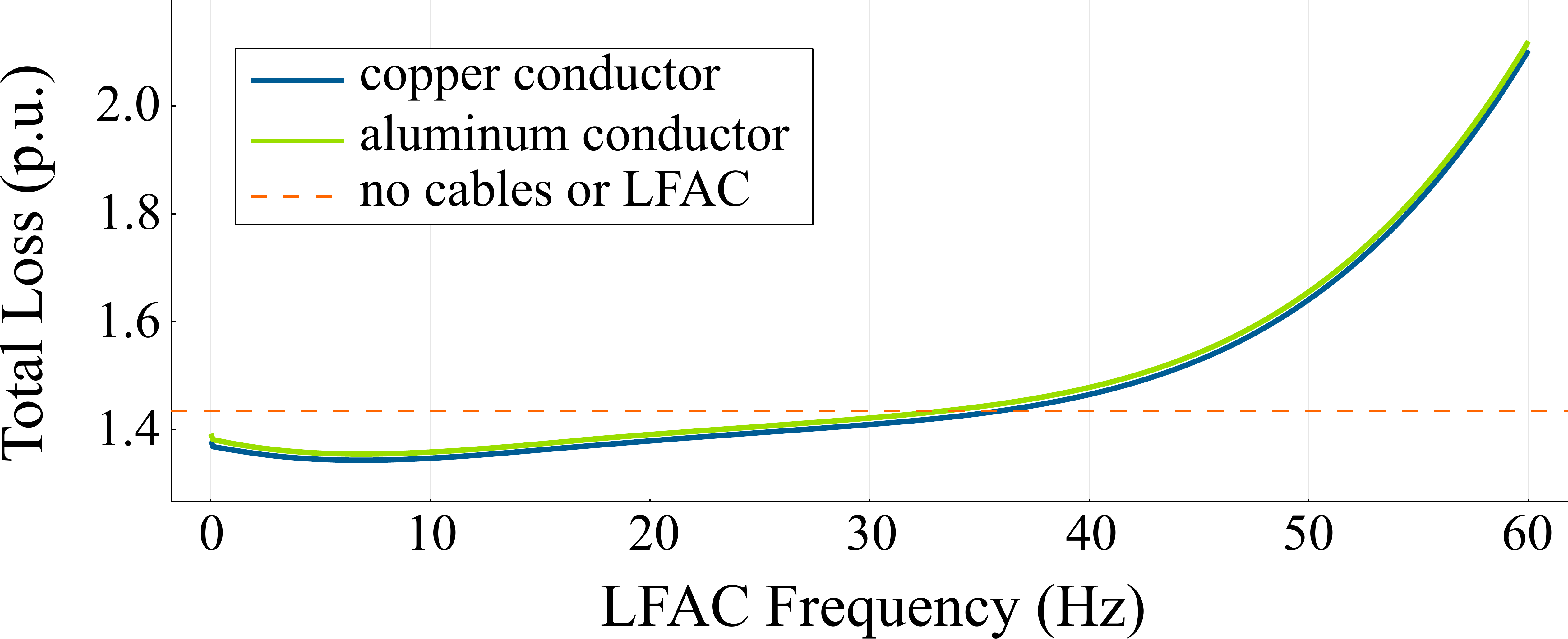}
\caption{Total losses in the network under scenario 1 for a sweep of LFAC frequencies, including a comparison of equivalent copper and aluminum conductors.}
\label{fig:sweeploss}
\end{figure}

Figure \ref{fig:sweeploss} plots the total losses in the network for the frequency sweep. The significant decrease in losses as frequency decreases matches the frequency dependence of resistance and conductance shown in Fig. \ref{fig:fit}. Much of the benefit of lowering the frequency to minimize losses can be gained by slight reductions of frequency.
We conclude that operating cables at a low frequency allows substantial loss reduction.

\subsubsection{Intra-Area Scenario}
For the intra-area undergrounding scenario, we again solve the problem without converters, with frequency as a variable, with frequencies fixed at 60 Hz and 16.7 Hz, and with DC cables. The results are listed in the second column of Table \ref{tab:comparison}. The solution without LFAC converters leads to a solution of 238.36$\cdot10^3$ USD, similar to the value before undergrounding. All of the solutions with LFAC converters reduce the cost by approximately 2.9\% to around 231.35$\cdot10^3$, demonstrating the value of power flow control with the converters. In contrast to scenario 1, however, the improvements do not change significantly with frequency, which is likely because the cable network lies within a single area and involves a large number of generators, which helps mitigate congestion and allows absorption of reactive power by the generators.

\section{Conclusion}\label{sec:conclusion}
This paper introduces a frequency-dependent model for underground cables which is accurate, yet tractable for inclusion in an optimal power flow model. This model utilizes detailed representations of the interactions between conductors and sheaths for three phase cables and fits approximate polynomials to represent the positive sequence series impedances and shunt admittances. The model is incorporated into a variable frequency optimal power flow formulation to enable the analysis of LFAC cables in system operation. %

We analyze LFAC cables in a system where overhead lines have been converted to underground cables to mitigate the risk of wildfire ignition.
Our results demonstrate that the active and reactive power control lead to large improvements, even enabling feasible operation where a feasible point was not previously found. In addition, cable losses depend significantly on frequency, with lower frequencies producing lower losses. We also observe that the angle limits in combination with large reactive power flows can limit transfer capacity on long, heavily loaded cables. %

Tools for quantification of these factors are critical to analyze the potential application of LFAC to a system with underground cables. When the operational benefits are substantial, they may be shown with these methods applied over a range of operating points to give lifetime system-level benefits which outweigh the capital cost of the converter stations. Combining these operational modeling and optimization tools with a detailed economic analysis is an important future area of work for low frequency AC transmission.

The applications of the cable model and associated optimal power flow model enable analysis of LFAC technology in a range of undergrounding applications, ranging from  %
system reliability under natural hazards to the design of economical offshore wind networks. Other future directions of model development include economic modeling of the frequency converters for LFAC and optimal placement of cables and converters for undergrounding decisions. %

\section*{Acknowledgment}
The authors thank Giri Venkataramanan at University of Wisconsin-Madison for the discussions that helped improve this work. D.K.S thanks the Decision and Infrastructure Sciences Division at Argonne National Laboratory for the support to present this work.
\bibliographystyle{IEEEtran}
\bibliography{IEEEabrv,refs}

\end{document}